\documentclass[useAMS,usenatbib]{mn2e}
\usepackage{graphicx}
\usepackage{epstopdf}
\usepackage{amsmath}
\usepackage{amssymb}

\title[Wind from hot accretion flow]{Two-dimensional inflow-wind solution of black hole accretion with an evenly symmetric magnetic field}
\author[Mosallanezhad, Bu \& Yuan]{Amin Mosallanezhad$^{1,2}$, Defu Bu$^{1}$ and Feng Yuan$^{1}$\thanks{E-mail: fyuan@shao.ac.cn (FY)}  \\
$^{1}$ Shanghai Astronomical Observatory, Chinese Academy of Sciences, 80 Nandan Road, Shanghai 200030, China\\
$^{2}$ University of Chinese Academy of Sciences, 19A Yuquan Road, Beijing 100049, China}
\begin{document}


\pagerange{\pageref{firstpage}--\pageref{lastpage}} \pubyear{2002}

\maketitle

\label{firstpage}
\begin{abstract}
We solve the two-dimensional magnetohydrodynamic (MHD) equations of black hole accretion with the presence of magnetic field. The field includes  a turbulent component, whose role is represented by the viscosity, and a large-scale ordered component. The latter is further assumed to be evenly symmetric with the equatorial plane. The  equations  are solved in the $r-\theta$ plane of a spherical coordinate by assuming time-steady and radially self-similar. An inflow-wind solution is found. Around the equatorial plane, the gas is inflowing; while above and below the equatorial plane at a certain critical $\theta$ angle, $\theta\sim 47^{\circ}$, the inflow changes its direction of radial motion and becomes wind.  The driving forces are analyzed and found to be the centrifugal force and the gradient of gas and magnetic pressure. The properties of wind are also calculated. The specific angular momentum of wind is found to be significantly larger than that of inflow, thus wind can transfer angular momentum outward. These analytical results are compared to those obtained by the trajectory analysis based on MHD numerical simulation data and good agreements are found.
\end{abstract}

\section{Introduction}

Black hole accretion models can be divided into hot and cold types
according to the temperature of the accretion flow. Since the
pioneer work of \citet{Narayan and Yi 1994} (see also
\citealt{Ichimaru 1977, Rees et al. 1982}), intensive studies have
been performed on black hole hot accretion flows. It is now known
that hot accretion flow is very common in the universe, ranging from
low-luminosity active galactic nuclei (LLAGNs), which is the
majority of nearby galaxies, to the quiescent and hard states of
black hole X-ray binaries. \citet{Yuan and Narayan 2014} have
recently reviewed our current understanding of hot accretion flows,
including its one-dimensional and multi-dimensional dynamics, jet
formation, radiation, and various astrophysical applications.

One of the most interesting topics in recent years in this field is
wind (also often called outflow). Wind is not only a fundamental
aspect of accretion flow, but also plays an important role in AGN
feedback, since wind can suppress the star formation and black hole
growth effectively (e.g., \citealt{Ostriker et al. 2010}). The study
of wind was initiated by the hydrodynamical and MHD numerical
simulations of accretion flows (\citealt{Stone Pringle and Begelman
1999}; \citealt{Igumenshchev and Abramowicz 1999}, 2000;
\citealt{Stone and Pringle 2001}). These simulations found that in
contrast to our original picture, the mass accretion rate decreases
with decreasing radius (see review in Yuan et al. 2012b). To explain
such an astonishing result, two models have been proposed, namely
convection-dominated accretion flow (CDAF; \citealt{Narayan et al.
2000}; \citealt{Quataert and Gruzinov 2000}) and adiabatic
inflow-outflow solution (ADIOS; \citealt{Blandford and Begelman
1999}). By comparing the  properties of inflow and outflow and
studying the convective stability of MHD accretion flow, Yuan et al.
(2012a; see also \citealt{Li Ostriker and Sunyaev 2013}) show that
strong wind must exist and it is the wind rather than convection
that results in the inward decrease of accretion
rate\footnote{\citet{Narayan et al. 2012} also studied the outflow
in hot accretion flow. While they also find the existence of wind,
their wind is much weaker than that found by Yuan et al. 2012a. The
reason for the discrepancy was analyzed in \citet{Yuan et
al.2015}.}. Most recently, based on MHD simulation data, \citet{Yuan
et al.2015} have convincingly show the existence of strong wind by
obtaining the trajectory of the virtual test particles based on 3D
GRMHD simulation data of black hole accretion\footnote{Bu et al.
(2013) show that the wind becomes much weaker if the angular
momentum of the accretion flow is very low.}. Moreover, the driving
forces of wind have also been analyzed and the detailed properties
of winds have been calculated, such as the mass flux, angular
distribution, terminal velocity, and fluxes of energy and momentum.
The theoretical study on the existence of strong wind has been
confirmed by the detection of emission lines from the accretion flow
around the supermassive black hole in the Galactic center, Sgr A*
(\citealt{Wang et al. 2013}). As an application of the accretion
wind theory, more recently, the formation of the Fermi bubbles
detected by the {\it Fermi} telescope in the Galaxy has  been
successfully explained by the interaction of wind launched from the
accretion flow around Sgr A* and the interstellar medium
(\citealt{Mou et al. 2014}; 2015). Compared to other theoretical
models of the Fermi bubbles, this ``accretion-wind'' model has two
advantages. One is that the parameters of wind adopted in the model
are taken from the MHD simulation of accretion flow. The second one
is that the model can explain the new observational results obtained
by {\it Suzaku} and {\it Chandra} which are in conflict with the
predictions of other models.

Almost all of the above-mentioned theoretical works on wind are
numerical. Compared to numerical simulations, analytical study has
its advantage of more clearly revealing some underlying physics. In
this paper, using analytical method, we solve the axisymmetric
steady solution of hot accretion flow with the presence of magnetic
field. Self-similar assumption is adopted as in many previous works.
Special attention is paid to whether the winds exist in the solution
and what are their driving mechanisms if they exist. The aim is to
compare with the numerical results presented in \citet{Yuan et
al.2015} thus to improve our understanding of numerical studies.

There have been many analytical works based on the self-similar
assumption. In the one-dimensional works, the existence of wind is
often an assumption (e.g. \citealt {Blandford and Begelman 1999};
\citealt{Akizuki and Fukue 2006}; \citealt{Abbassi et al. 2008};
\citealt{Zhang and Dai 2008}; \citealt{Bu et al. 2009}). Winds have
been found in some two-dimensional solutions (e.g. \citealt{Xu and
Chen 1997}; \citealt{Blandford and Begelman 2004}; \citealt{Xue and
Wang 2005}; \citealt{Tanaka and Menou 2006}; \citealt{Jiao and Wu
2011}; Mosallanezhad et al. 2014; Gu 2015; Samadi \& Abbassi 2016 )
but not all (e.g., \citealt{Narayan and Yi 1995} and references
therein). In all these works magnetic field is not included. It is
well known that magnetic field must be present and even plays an
important role in the dynamics of accretion flow, such as the
angular momentum transfer by MRI (\citealt{Balbus and Hawley 1998}),
the convective stability of accretion flow, which is related with
the wind production (Yuan et al. 2012a; \citealt{Narayan et al.
2012}), and the driving mechanism of winds (\citealt{Yuan et
al.2015}). So it is useful to solve the accretion solutions by
including magnetic field. This is the aim of the present paper.

The organization of the paper  is as follows. In \S2, the
basic MHD equations and assumptions are given. The numerical solutions of the equations are described in \S3. The results are summarized in  \S4.

\section{Basic MHD equations }

The basic MHD equations describing accretion flows read:
\begin{gather}
    \frac{d \rho}{d t} + \rho \mathbf{\nabla} \cdot \mathbf{v} = 0  \label{continuity_main},\\
    \rho  \frac{d \mathbf{v}}{d t}  = -\rho \mathbf{\nabla}\Psi - \mathbf{\nabla}p + \nabla \cdot \mathbf{T} + \frac{1}{c} \left( \mathbf{J}\times \mathbf{B} \right)   \label{motion_main},\\
    \rho \frac{d e}{dt} - \frac{p}{\rho} \frac{d \rho}{dt}  =  f Q^{+}  \label{energy_main},\\
    \frac{\partial \mathbf{B}}{\partial t} = \mathbf{\nabla}\times \left( \mathbf{v}\times \mathbf{B} - \frac{4 \pi}{c}\eta \mathbf{J}  \right) \label{induction_main},\\
    \mathbf{\nabla} \cdot \mathbf{B} = 0 \label{divb}.
\end{gather}
In the above equations, $ \rho $, $ \mathbf{v} $, $ p $, $e$, $
\mathbf{T} $, $ \mathbf{B} $ and $ \eta $ denote the
density, velocity, gas pressure, gas internal energy, viscous stress
tensor, magnetic field, and magnetic diffusivity, respectively. We adopt an
equation of state of ideal gas, $p=(\gamma-1)\rho e$, where $\gamma$
is the specific heat ratio. A spherical coordinate ($r$, $\theta$, $\phi$) is adopted in the present work. $ \Psi (= - GM/r) $ is the Newtonian
gravitational potential, where $M$ is the mass of the central black hole,
$G$ is the gravitational constant. $ \mathbf{J} \equiv
c/4\pi(\mathbf{\nabla} \times \mathbf{B}) $ is the electric current. We must include this dissipation term in eq. (4), because otherwise we can't obtain the steady solution due to the accumulation of magnetic flux in the accretion.
$ Q^+ $ is the heating rate. $ f $ is the advection factor describing
the fraction of the heating energy which is stored in the gas and
advected into the black hole. In this paper, we set $ f = 1 $. The
heating rate $ Q^+ $ can be decomposed into two components,
\begin{equation} \label{heating rate}
 Q^{+} = Q_{vis} + Q_{res},
\end{equation}
with $ Q_{vis} $ and $ Q_{res} $ denoting the viscous heating and the
magnetic field dissipation heating, respectively. Following Stone et al. (1999), we assume that the
only non-zero component of the viscous tensor $ \mathbf{T} $ is the
azimuthal component,
\begin{equation} \label{T_rphi}
 T_{r\phi} = \rho \nu r \frac{\partial (v_{\phi}/r)}{\partial r},
\end{equation}
where $ \nu $ is the kinematic viscosity. The viscous heating and
the magnetic field dissipation heating can be described as,
\begin{gather}
  Q_{vis} = T_{r\phi} r \frac{\partial}{\partial r} \left( \frac{v_{\phi}}{r} \right), \label{Q_vis}\\
  Q_{res} = \frac{4 \pi }{c^{2}} \eta \mathbf{J}^2.\label{Q_res}
\end{gather}
We adopt the usual $ \alpha $ description of viscosity
$ \nu= \alpha p/ (\rho \Omega_{K}) $, where $ \Omega_{K} =
(GM/r^{3})^{1/2} $ is the Keplerian angular velocity. In order to satisfy
the radial self-similar condition we assume the magnetic diffusivity
$ \eta = \eta_{0} p/ (\rho \Omega_{K}) $.

Following the suggestion from numerical MHD simulations, we
decompose the magnetic field into a large-scale component and a
turbulent component. Both of them can transfer the angular momentum
and can be dissipated and produce heat.  The magnetic field $
\mathbf{B} $ in equations (\ref{motion_main}),
(\ref{induction_main}) and (\ref{divb}) corresponds to the former.
We describe the effects of the turbulent component of the magnetic
field in transferring the angular momentum and in dissipating the
energy through the usual $ \alpha $ description. Specifically, the
viscous force $ \nabla \cdot \mathbf{T} $ in equation
(\ref{motion_main}) represents angular momentum transfer by the
turbulent magnetic field; the viscous heating rate $ Q_{vis} $ in
equation (\ref{heating rate}) is correspondingly associated with the
turbulent component of the magnetic field.

We assume that the large-scale magnetic field is evenly symmetric about
the equatorial plane (see Figure 1 for the poloidal configuration of the
magnetic field). This kind of symmetry is widely adopted in the studies
of accretion disks (e.g. \citealt{Blandford and Payne 1982};
\citealt{Lovelace et al. 1994}; 
\citealt{Cao 2011}; \citealt{Li and Begelman 2014})\footnote{Shahram Abassi's group is working on the case of an oddly symmetric magnetic field at the same time with us.}. Thus we have,
\begin{gather}
  B_{r}(r, \theta) = - B_{r}(r, \pi - \theta), \label{even symmetry r} \\
  B_{\theta}(r, \theta) = + B_{\theta}(r, \pi - \theta), \label{even symmetry th}\\
  B_{\phi}(r, \theta) = - B_{\phi}(r, \pi - \theta), \label{even symmetry phi}
\end{gather}
A toroidal component of the magnetic field is generated due to the
shear of the accretion flow. Therefore the radial and the toroidal
components of the magnetic field should have opposite sign. In this
paper, we only focus on the region above the equatorial plane, we
set the radial component of themagnetic field to be positive ($ B_r
> 0 $). Therefore, the toroidal component of the magnetic field
generated by shear is negative ($ B_{\phi} < 0 $).

We solve for the steady state, axisymmetric  ($\partial/\partial
t=\partial/\partial \phi=0$) solutions of eqs. 1-5. The continuity
equation can be rewritten as:
\begin{equation}\label{continuty_ref}
\frac{1}{r^{2}}\frac{\partial}{\partial r}\left(r^{2}\rho v_{r}\right) + \frac{1}{r \sin\theta}\frac{\partial}{\partial \theta}\left(\sin\theta \rho v_{\theta}\right) = 0,
\end{equation}
The momentum equation (\ref{motion_main}) reads:

\begin{multline}\label{motionr-ref}
\rho \left[ v_{r} \frac{\partial v_{r}}{\partial r}+\frac{v_{\theta}}{r} \left(\frac{\partial v_{r}}{\partial \theta} - v_{\theta}\right)- \frac{v_{\phi}^{2}}{r} \right] =
- \rho \frac{GM_{*}}{r^{2}} - \frac{\partial p}{\partial r} \\
+ \frac{1}{4\pi} \left( J_{\theta} B_{\phi} - J_{\phi}B_{\theta} \right),
\end{multline}

\begin{multline}\label{motiont_ref}
\rho \left[ v_{r} \frac{\partial v_{\theta}}{\partial r} + \frac{v_{\theta}}{r}\left(\frac{\partial v_{\theta}}{\partial \theta} + v_{r}\right) - \frac{v_{\phi}^{2}}{r}\cot \theta \right] = -\frac{1}{r} \frac{\partial p}{\partial \theta} \\
+ \frac{1}{4\pi} \left( J_{\phi} B_{r} - J_{r}B_{\phi} \right),
\end{multline}

\begin{multline}\label{motionp_ref}
\rho \left[ v_{r} \frac{\partial v_{\phi}}{\partial r} + \frac{v_{\theta}}{r} \frac{\partial v_{\phi}}{\partial \theta} + \frac{v_{\phi}}{r} \left(v_{r} + v_{\theta} \cot \theta \right) \right] = \frac{1}{ r^{3}} \frac{\partial}{\partial r} \left( r^{3} T_{r \phi} \right)\\
+ \frac{1}{4\pi} \left( J_{r} B_{\theta} - J_{\theta}B_{r} \right),
\end{multline} where the current  ($ \mathbf{J} $) reads:
\begin{gather}\label{jr}
  J_{r} = \frac{1}{r \sin\theta} \frac{\partial}{\partial \theta} \left( B_{\phi} \sin\theta \right),\\
  J_{\theta} = - \frac{1}{r} \frac{\partial}{\partial r} \left( r B_{\phi} \right),\\
  J_{\phi} = \frac{1}{r} \left[ \frac{\partial}{\partial r} \left( r B_{\theta} \right) - \frac{\partial B_{r}}{\partial \theta} \right].
 \end{gather}
The equation of energy is expressed as:
 \begin{multline}\label{energy1}
  \rho \left( v_{r} \frac{\partial e}{\partial r} + \frac{v_{\theta}}{r} \frac{\partial e}{\partial \theta} \right) - \frac{p}{\rho} \left( v_{r} \frac{\partial \rho}{\partial r} + \frac{v_{\theta}}{r} \frac{\partial \rho}{\partial \theta} \right) = \\
  f \left( T_{r\phi} r \frac{\partial}{\partial r} \left( \frac{v_{\phi}}{r} \right) + \frac{\eta}{4\pi} \left( J_{r}^{2} + J_{\theta}^{2} + J_{\phi}^{2} \right) \right).
\end{multline}
The three components of induction equation (4) can be expressed as:
\begin{gather}
\frac{\partial B_r}{\partial t}=\frac{\partial}{\partial \theta}
\left[ r \sin\theta \left( v_{r} B_{\theta} - v_{\theta} B_{r}  -
\eta J_{\phi} \right) \right], \label{induction1_ref}
\end{gather}

\begin{gather}
\frac{\partial B_\theta}{\partial t}=\frac{\partial}{\partial r}
\left[ r \sin\theta \left( v_{r} B_{\theta} - v_{\theta} B_{r} -
\eta J_{\phi} \right) \right], \label{induction2_ref}
\end{gather}

\begin{multline}
\frac{\partial B_\phi}{\partial t}=\frac{\partial}{\partial r} \left( r v_{\phi} B_{r} - r v_{r} B_{\phi} \right) + \frac{\partial}{ \partial \theta} \left( v_{\phi} B_{\theta} - v_{\theta} B_{\phi} \right) + \frac{\partial}{\partial \theta} \left(\eta J_{r}\right) \\
- \frac{\partial}{\partial r} \left(r \eta
J_{\theta}\right).\label{induction3_ref}
\end{multline}

For a steady state, the right hand side of equations
(\ref{induction1_ref}), (\ref{induction2_ref}) and
(\ref{induction3_ref}) has to be equal to zero. So we have $ r
\sin\theta \left( v_{r} B_{\theta} - v_{\theta} B_{r} - \eta
J_{\phi} \right)=0 $ or a constant.  In this paper, we assume it is
equal to zero:
\begin{equation}\label{induction-r-theta}
r \sin\theta \left( v_{r} B_{\theta} - v_{\theta} B_{r}  - \eta J_{\phi} \right)=0.
\end{equation}
Finally, the divergence-free equation (\ref{divb}) can be written
as,\begin{gather}
  \frac{1}{r^{2}} \frac{\partial}{\partial r} \left( r^{2} B_{r} \right) + \frac{1}{r \sin\theta} \frac{\partial}{\partial \theta} \left( \sin\theta B_{\theta} \right) = 0. \label{DivB}
\end{gather}


Similar to previous works (e.g., \citealt{Narayan and Yi 1995}; \citealt{Xue and Wang 2005}; \citealt{Jiao and Wu 2011}), we seek the radially self-similar solutions in the following forms:
\begin{gather}
    \rho (r, \theta) = \rho(\theta) r^{-n}, \label{self-rho} \\
    v_{r} (r, \theta ) = \sqrt{\frac{GM}{r}} v_{r} (\theta)  = r \Omega_{K}(r)v_{r}(\theta), \label{self-vr} \\
    v_{\theta}(r, \theta) = r \Omega_{K}(r) v_{\theta}(\theta),  \label{self-vtehta} \\
    v_{\phi} (r, \theta) = r \Omega_{K}(r) v_{\phi}(\theta),  \label{self-vphi} \\
    p (r, \theta) = p(\theta) GM r^{-n-1} \label{self-p}, \\
    B_{r} (r, \theta) = b_{r} (\theta) \sqrt{GM} r^{-(n/2)- (1/2)}, \label{self-Br}\\
    B_{\theta} (r, \theta) = b_{\theta} (\theta) \sqrt{GM} r^{-(n/2)- (1/2)}, \label{self-Bt}\\
    B_{\phi} (r, \theta) = b_{\phi}(\theta) \sqrt{GM} r^{-(n/2)- (1/2)}. \label{self-Bp}
\end{gather}

\begin{figure*}
\begin{center}
\includegraphics[width=100mm, angle=0]{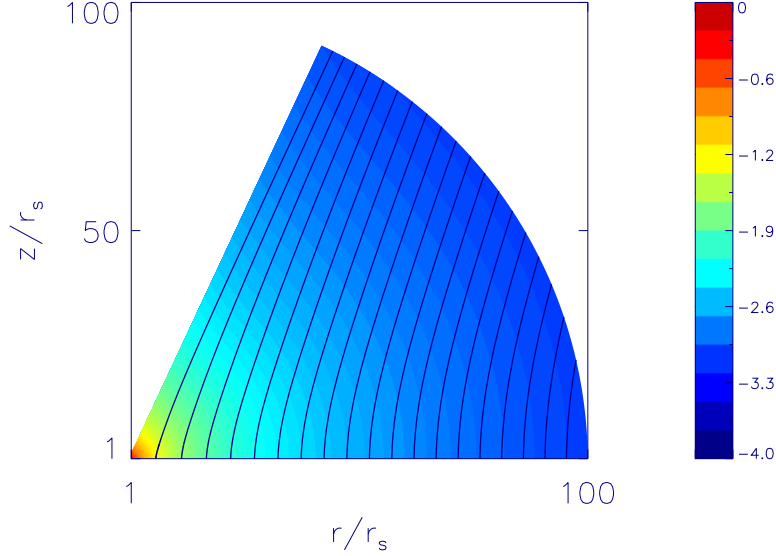}
\vspace{1cm} \caption{Solid lines represent the magnetic field line.
Colors show the logarithm angular velocity ($v_\phi/r$). It is clear
that moving along the magnetic field lines starting from the
equatorial plane, the angular velocity decreases. Therefore, if we
assume that $B_r$ above the equatorial plane is positive, shear of
the flow will produce a negative $B_\phi$. } \label{magneticfield}
\end{center}
\end{figure*}

Using the above self-similar assumptions, equations
(\ref{continuty_ref})-(\ref{motionp_ref}), (\ref{energy1}),
(\ref{induction3_ref})-(\ref{DivB}) can be written as,
\begin{equation}\label{rho_new}
v_{\theta} \frac{d \ln \rho}{d\theta} +  \left( \frac{3}{2} - n \right) v_{r} + v_{\theta} \cot\theta  + \frac{dv_{\theta}}{d \theta}  = 0,
\end{equation}
\begin{multline} \label{vr_new}
 \rho \left[ - \frac{1}{2} v^{2}_{r} +  v_{\theta} \frac{d v_{r}}{d\theta}  -  v^{2}_{\theta} - v^{2}_{\phi} \right] = - \rho + (n + 1)p \\
 + \frac{1}{4\pi} \left( j_{\theta} b_{\phi} - j_{\phi}b_{\theta} \right),
\end{multline}
\begin{multline}\label{vt_new}
 \rho \left[ \frac{1}{2} v_{r} v_{\theta} + v_{\theta} \frac{d v_{\theta}}{d\theta} - v_{\phi}^{2} \cot\theta \right] = - \frac{dp}{d\theta} \\
 + \frac{1}{4\pi} \left( j_{\phi}b_{r} - j_{r}b_{\phi} \right),
\end{multline}
\begin{multline}\label{vp_new}
  \rho \left[ \frac{1}{2} v_{r} v_{\phi} + v_{\theta} \frac{dv_{\phi}}{d\theta} + v_{\theta} v_{\phi} \cot\theta \right] =  \frac{3}{2}\alpha  (n - 2) p v_{\phi} \\
   + \frac{1}{4\pi} \left( j_{r}b_{\theta} - j_{\theta}b_{r} \right),
\end{multline}
\begin{multline}\label{p_new}
 (n\gamma - n - 1)v_{r}p + v_{\theta}\frac{dp}{d\theta} - \gamma p v_{\theta}  \frac{d \ln \rho}{d\theta} = \\
f \left( \gamma - 1 \right)  \left(  \frac{9}{4} \alpha p v_{\phi}^{2} + \frac{\eta}{4\pi} \left( j_{r}^{2} + j_{\theta}^{2} + j_{\phi}^{2} \right) \right),
\end{multline}
\begin{multline}\label{induction3_new}
  \frac{n}{2} \left( v_{r} b_{\phi} - v_{\phi} b_{r}  + \eta j_{\theta} \right) + v_{\phi} \frac{db_{\theta}}{d\theta} + b_{\theta} \frac{dv_{\phi}}{d\theta} - v_{\theta} \frac{db_{\phi}}{d\theta} - b_{\phi} \frac{dv_{\theta}}{d\theta} \\
  + \eta \frac{d j_{r}}{d \theta} + j_{r} \frac{d \eta}{d \theta} = 0,
\end{multline}
\begin{equation}\label{divb_new}
  \frac{db_{\theta}}{d\theta} - \frac{1}{2}(n - 3)b_{r} + b_{\theta}\cot\theta = 0.
\end{equation}
where
\begin{equation}\label{eta}
  \eta = \eta_{0} \frac{p}{\rho},
\end{equation}
\begin{gather}\label{js1}
    j_{r} = \frac{db_{\phi}}{d\theta} + b_{\phi} \cot\theta, \\
    j_{\theta} = \frac{1}{2} (n - 1)b_{\phi}, \\
    j_{\phi} = \frac{1}{\eta } \left( v_{r} b_{\theta} - v_{\theta} b_{r} \right).
\end{gather}
Equations (\ref{rho_new})-(\ref{divb_new}) are differential
equations for eight variables: $ v_{r}(\theta) $, $
v_{\theta}(\theta) $, $ v_{\phi}(\theta) $, $ \rho(\theta) $, $
p(\theta) $, $ b_{r}(\theta) $, $ b_{\theta}(\theta) $ and $
b_{\phi}(\theta) $. In this paper, we assume the accretion flow is
evenly symmetric about the equatorial plane, then $ \rho(\theta) =
\rho(\pi - \theta) $, $ p(\theta) = p(\pi - \theta) $, $
v_{r}(\theta) = v_{r}(\pi - \theta) $, $ v_{\phi}(\theta) =
v_{\phi}(\pi - \theta) $ and $ v_{\theta}(\theta) = - v_{\theta}(\pi
- \theta) $. At the equatorial plane, we have by symmetry:
\begin{equation}\label{boundry conditions}
\theta =\frac{\pi}{2} :  b_{r} = v_{\theta} = \frac{d v_{r}}{d
\theta} = \frac{d v_{\phi}}{d \theta} = \frac{d \rho}{d \theta} =
\frac{d p}{d \theta} =  \frac{d b_{\theta}}{d \theta} = \frac{d
b_{\phi}}{d \theta} = 0.
\end{equation}
We set the density at the equatorial plane $ \rho(\pi/2) = 1 $.
Under the above boundary conditions, equations
(\ref{rho_new})-(\ref{divb_new}) can be simplified into the
following equations,
\begin{equation}\label{bandary-1}
    \frac{d v_{\theta}}{d \theta} = \big( n - \frac{3}{2} \big) v_{r},
\end{equation}
\begin{equation}  \label{bandary-2}
    \frac{1}{2} v_{r}^{2} + v_{\phi}^{2} + \left(n + 1\right) p  +  \frac{1}{4\pi} \left( \frac{1}{2} \left( n - 1 \right) b_{\phi}^{2} + \frac{v_{r} b_{\theta}^{2}}{\eta_{0}p } \right)=0,
\end{equation}
\begin{equation}  \label{bandary-3}
  v_{r} = 3 \alpha \left( n - 2 \right) p,
\end{equation}
\begin{multline}  \label{bandary-4}
   \left( n \gamma - n - 1 \right)v_{r} = f \left( \gamma - 1 \right) \times \\
    \left[ \frac{9}{4} \alpha v_{\phi}^{2} + \frac{\eta_{0}}{4 \pi} \left(  \left( \frac{1}{2} \left( n - 1 \right) b_{\phi} \right)^{2} +  \left( \frac{v_{r} b_{\theta}}{\eta_{0}p} \right)^{2}  \right) \right].
\end{multline}
At the equatorial plane, we set the ratio of the magnetic pressure to the gas pressure as,
\begin{gather}
    \beta_{\theta0} =  \frac{b^{2}_{\theta}}{ 8 \pi p_{g}} \underset{\pi/2}{\bigg|}, \label{betatt}\\
    \beta_{\phi0} =   \frac{b^{2}_{\phi}}{ 8 \pi p_{g}} \underset{\pi/2}{\bigg|}. \label{betatp}
\end{gather}

In principle, to solve eqs. (\ref{rho_new})-(\ref{divb_new}), in addition to the boundary condition at $\theta=90^{\circ}$, we should also supply the boundary condition at the rotation axis, i.e., $\theta=0^{\circ}$. Mathematically, this is a two-point boundary problem and it will assure that we have a well-behaved solution in the whole $r-\theta$ space. We have tried to obtain such solutions but failed. Therefore, we only require the solution  satisfying  the boundary condition at $\theta=90^{\circ}$. As we will explain in the next section, we think the solution obtained in this way should still be physically meaningful.

\begin{figure*}
\begin{center}
\includegraphics[width=\textwidth, angle=0]{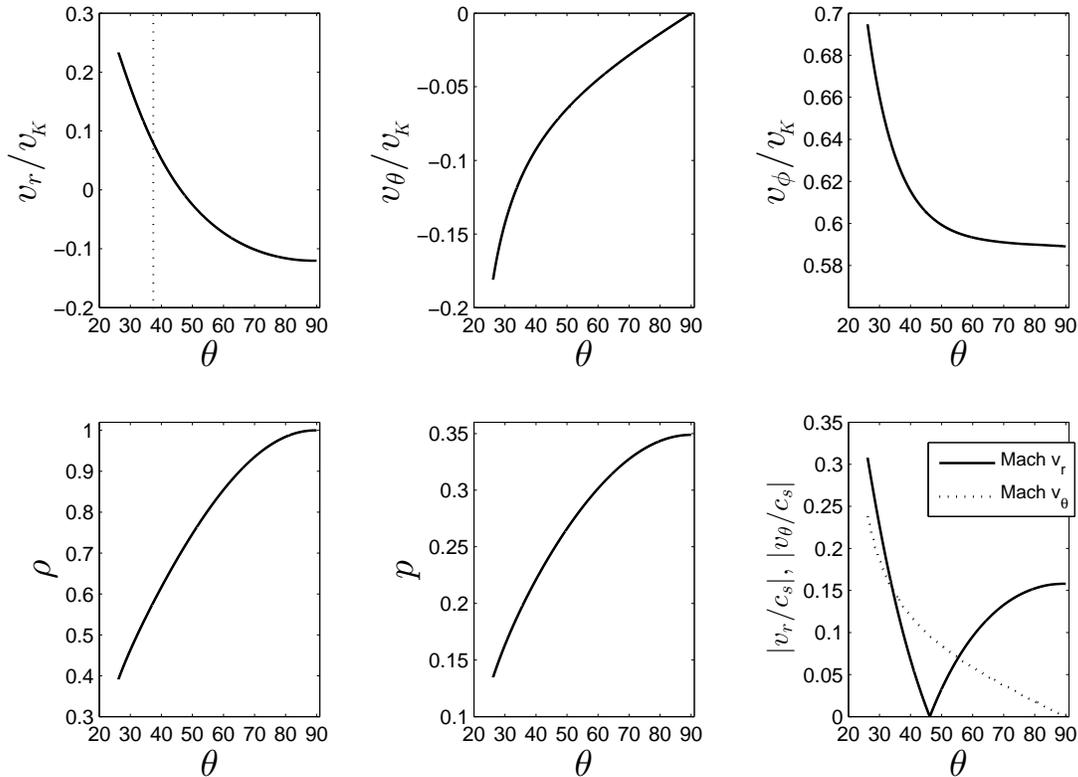}
\caption{Angular profile of a variety of variables. The up-left plot
shows that the radial velocity changes its sign at $\theta\sim
47^{\circ}$.} \label{velocites}
\end{center}
\end{figure*}

\begin{figure*}
\begin{center}
\includegraphics[width=125mm, angle=0]{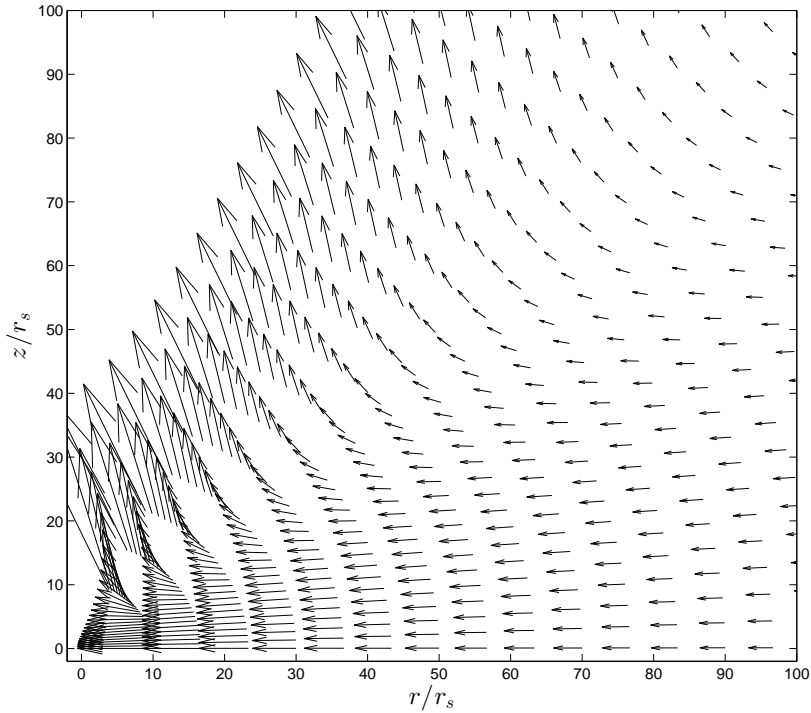}
\caption{Stream lines of  the accretion flow, showing the
inflow-wind structure. The length of arrows indicates the absolute
value of $\vec{v}_r+\vec{v}_\theta$.} \label{velocityfield}
\end{center}
\end{figure*}

\begin{figure*}
\begin{center}
\includegraphics[width=120mm, angle=0]{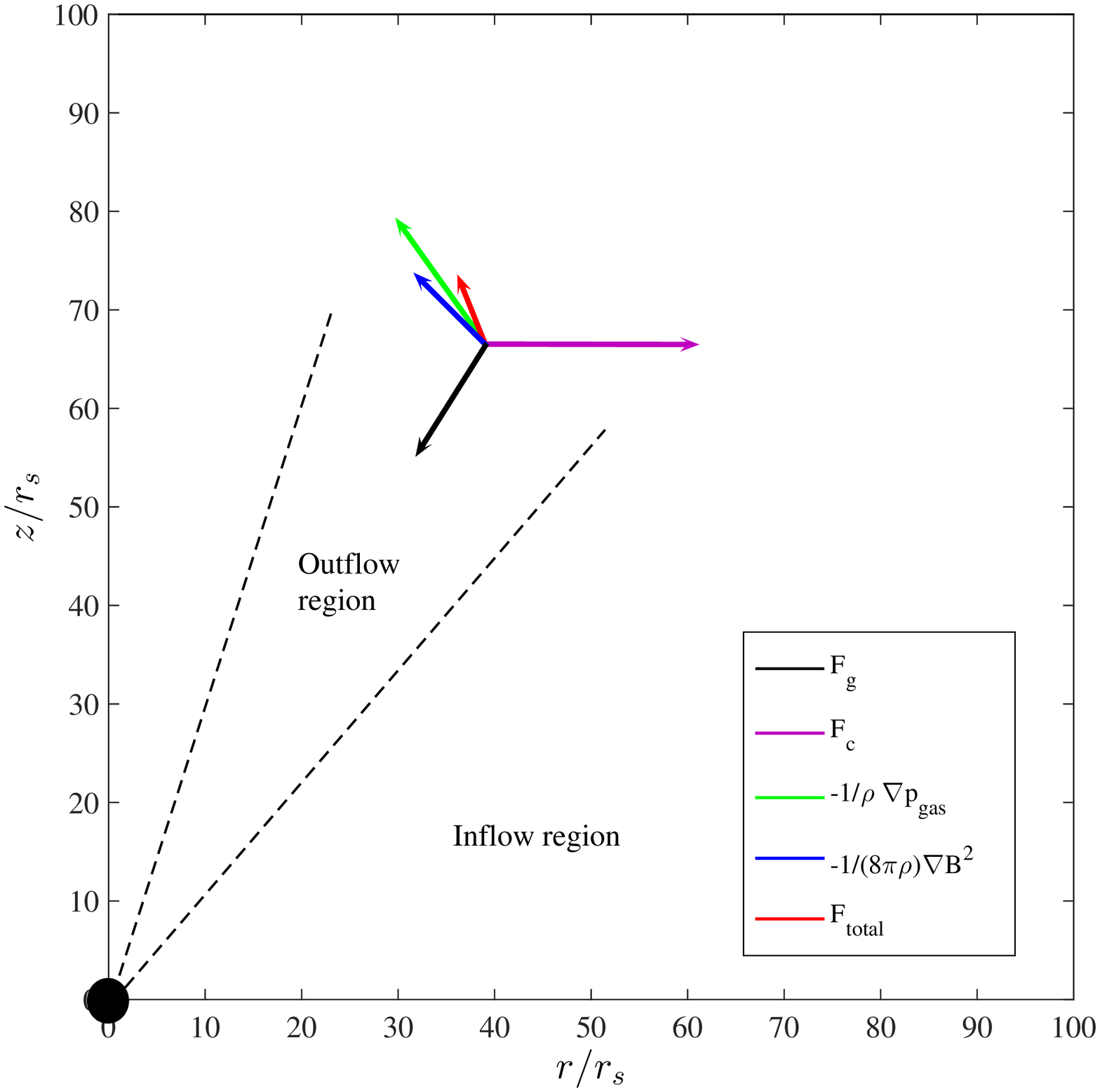}
\caption{The force analysis at the wind region to show the driving mechanism of wind. The length of the arrows
schematically denotes the magnitude of the forces while the
direction for the direction of forces. $F_g$: graviational force; $F_c$: centrifugal force; $F_{\rm total}$: the total force.}
\label{Schematic}
\end{center}
\end{figure*}

\section{Results}
We solve the above equations (\ref{rho_new})-(\ref{divb_new})
numerically. We integrate these equations from the equatorial plane
($ \theta=90^{\circ} $) towards the rotational axis
($\theta=0^{\circ} $).  We adopt the values of $\beta_{\theta0}$ and
$\beta_{\phi0}$ with reference to the numerical simulation data of
Yuan et al. (2012a), but our results are not sensitive to the exact
values. As we have explained before, the toroidal component of the
magnetic field, which is produced by shear, should be negative. We
do find that the toroidal component is negative at the beginning of
integration.  But at a certain critical value of $\theta$, denoted
as $\theta_B$, we find that it begins to change its sign. At the
region of  $\theta<\theta_B$, the toroidal component of the magnetic
field becomes positive thus the solution is  no longer physical so
we stop our integration at $\theta_B$.  We think our solution in the
region of $\theta_B<\theta<90^{\circ}$ is still
physical\footnote{The two-dimensional solution obtained by
\citet{Jiao and Wu 2011} is similar in the sense that they also have
to stop their integration at a certain $\theta>0^{\circ}$.}. This is
because, on the one hand, the solution satisfies the equations and
the boundary conditions at $\theta=90^{\circ}$. On the other hand,
the values of all physical quantities of the accretion flow at
$\theta=\theta_B$ are physical. Therefore, we can reasonably treat
these values as boundary conditions at $\theta_B$. In fact, as we
will describe below, the main properties of the solutions we obtain
in this way are in good agreement with those obtained in \citet{Yuan
et al.2015} from numerical simulations.

From the bottom-right plot of Figure 2, it seems that there is a
sonic point for $v_{\theta}/c_s$ and the integration stops when the
sonic point is approached because of the singularity there. This is
not true. This is because, the value of $v_{\theta}$ must be equal
to zero at the pole because of the axisymmetric boundary condition
there. Therefore, $v_{\theta}/c_s$ should not pass through a sonic
point. We note that Jiao \& Wu (2011) obtained a similar result with
us, i.e., their solution also has to stop at a certain $\theta$
angle. An interesting explanation for such a ``truncation'' is given
in their work.

\subsection{Overall inflow-wind structure of the solutions}
Figure \ref{magneticfield} shows the poloidal magnetic field configuration. We emphasize
that while the even symmetry of the field is our assumption, the strength of the field is obtained by solving equations (\ref{rho_new})-(\ref{divb_new}).  Above the
equatorial plane, the radial component of the magnetic field is
positive (pointing to large radii). So the toroidal component of the
magnetic field, which is produced by shear, is negative.  We find that such kind of magnetic field configuration is  similar to the time-averaged field configuration in MHD numerical simulation in Yuan et al. (2012a) and Yuan et al. (2015).

Figure \ref{velocites} presents the angular profiles of four
quantities. The parameters we adopt are  $ \alpha = \eta_{0}= 0.1,
\gamma = 5/3, \beta_{\theta0} = 2 \times 10^{-4}, \beta_{\phi0} = 2
\times 10^{-3}, n = 0.85 $ and $ f = 1 $. From the equatorial plane
to the rotation axis, the density and pressure decrease, while $v_r$
and $v_\phi $ increase.
 This is consistent with that in Jiao \& Wu (2011). The value of
$v_\theta$ is negative and its absolute value increases towards the
pole. The Mach number $v_{\theta}/c_s$ increases towards the pole,
this is also consistent with that found in Jiao \& Wu (2011).
However, we note that beyond $\theta_B$, $v_\theta$ have to decrease
towards the pole and eventually become zero at pole due to the
boundary condition there.

In the region of $ 47^{\circ} < \theta < 90^{\circ}$, the radial
velocity is negative. This is the inflow region. At the region of $
\theta < 47^{\circ}$, the radial velocity becomes positive, i.e.,
wind is evident. Figure \ref{velocityfield} shows the steam line of
the flow. This result is fully consistent with the results obtained
in Yuan et al. (2015; see also  \citealt{Narayan et al. 2012};
\citealt{Sadowski et al. 2013}). In those works, they also found
that inflow is present around the equatorial plane of the accretion
flow while wind is present in the polar region. The boundary between
inflow and wind obtained in the present paper is even quantitatively
similar to that obtained in \citet{Yuan et al.2015}. In both cases
the boundary is located at about $\theta\sim 50^{\circ}-60^{\circ}$.

\subsection{Angular momentum transfer by wind}
The up-right panel of Figure \ref{velocites} shows the rotation
velocity of the accretion flow. We find that it significantly
increases from the equatorial plane to the rotation axis, which
suggests that the specific angular momentum of wind may be larger
than that of inflow. To examine this point in more detail, we have
calculated the mass flux-weighted specific angular momentum of both
the wind and inflow (see Eqs. 8 \& 9 in Yuan et al. (2012a) for the
details of calculation). In unit of the Keplerian angular momentum
$l_K$, the mass-flux weighted angular momentum of wind and inflow is
found to be
\begin{equation} \label{mass-flux weighted angular momentum}
l_{\rm wind} = 0.67 l_{K}, ~~~~ l_{\rm inflow}=0.59 l_{K},
\end {equation}
respectively. This result means that wind can transfer the angular
momentum of the accretion flow. This result is again consistent with
the numerical simulation result presented in Yuan et al. (2012a).

To compare the efficiency of angular momentum transfer by outflow
and by viscosity, we  calculate the angular momentum flux
transported by outflow and viscosity as follows \begin{equation}
 {F}_{L\rm out} = 2\pi r^{2} \int_{\theta_B}^{\pi/2} \rho \max(v_{r},0) (r v_\phi)
   \sin \theta d\theta,
   \label{inflowrate}
\end{equation}
\begin{equation}
 {F}_{L\rm vis} = 2\pi r^{2} \int_{\theta_B}^{\pi/2} r T_{r\phi}
   \sin \theta d\theta.
   \label{flowvisc}
\end{equation}
We find \begin{equation} \frac{F_{L\rm out}}{F_{L\rm vis}}=0.26.
\end{equation}
This means outflows do play an important role in transferring
angular momentum.

\subsection{Poloidal speed of wind}
We have also calculated the mass flux-weighted poloidal speed $ v_{p}(\equiv \sqrt{ v_{r}^{2} + v_{\theta}^{2} } $) of winds when they are just launched. The result is
\begin{equation} \label{vp}
v_{p}(r) = 0.18\ v_{K} (r),
\end{equation}
where $ v_{K} $ is the Keplerian velocity. This is
consistent with that obtained in Yuan et al. (2015),
where they found that $ v_{p} \approx 0.2 v_{K} (r) $.

\subsection{Bernoulli parameter and terminal velocity of wind}

The Bernoulli parameter is defined as
\begin {equation} \label{Bernoulli}
Be (r) = \frac{v^{2}}{2} + \frac{\gamma p}{(\gamma - 1) \rho} - \frac{GM}{r}.
\end{equation}
The first, second and the third terms on the right-hand side of the equation correspond to the kinetic energy, enthalpy and
gravitational energy, respectively.
We have calculated the mass flux-weighted Bernoulli parameter of the wind and obtained
\begin{equation} \label{Bernoulli1}
Be (r) = 0.11\ v_{K}^{2}(r).
\end{equation}
The value of $Be$ is positive, which means that the gas has enough energy to overcome the gravitational
potential, to do work to its surroundings, and to escape to infinity. As a comparison, the numerical simulation result obtained in Yuan et al. (2012a; eq. 19) is that $Be(r)/v_{K}^2(r)\approx (0.1-0.2)$. So the present result is again in good agreement with that obtained by numerical simulations\footnote{Yuan et al. (2015) find that $Be(r)$ is roughly constant of radius; but note that the constant $Be$ is along the trajectories of test particles.}.

Now, as in Yuan et al. (2012a), we try to estimate the terminal poloidal velocity of wind based on $Be(r)$. To do this, we assume that, once launched, the viscous stress in the wind can be ignored and $ Be (r) $ is conserved. When
$ r $ is large enough, the last two terms in Equation
(\ref{Bernoulli}) vanish. If the angular momentum is conserved (the wind is inviscid), the
rotational velocity is zero at large radius, so the terminal velocity is mainly the
polodial velocity. Therefore the terminal velocity is
\begin{equation} \label{terminal velocity}
v_{p,term}\approx v_p\approx \sqrt{2Be(r)}
\end{equation}
Combing with eq. (\ref{Bernoulli1}), we have
\begin{equation} \label{terminal velocity2}
v_{p,term} \approx 0.47\ v_{K} (r).
\end{equation}
So the terminal poloidal velocity of wind originating from radius $r$ is about half of the Keplerian velocity at their origin. As a comparison, Yuan et al. (2012a) find $v_{\rm p,term}(r)\approx 0.5v_{K}(r)$ while Yuan et al. (2015) find $v_{\rm p,term}(r)\sim (0.2-0.4)v_K(r)$.

\subsection{Analysis of forces driving the wind}
To study the driving mechanisms of wind, we have calculated the
forces at the wind region. Figure \ref{Schematic} shows the result.
We can see from the figure that the dominant driving forces are the
centrifugal force, the gradient of the gas pressure and the magnetic
pressure. The strength of the three forces are comparable. This
result is again fully consistent with that found in Yuan et al.
(2015), although here we have an ordered large-scale magnetic field
while the field in Yuan et al. (2015) is tangled. Jiao \& Wu (2011)
found that the driving forces are the gas pressure gradient and
centrifugal force because they do not include magnetic field in
their model. Compared to the Blandford \& Payne (1982) model, it is
similar in the sense that the centrifugal force is important; the
difference is that here the gas pressure and magnetic pressure
gradient forces play an equally important role in driving the winds.

\section{Summary and discussion}

In this paper, we have solved the two-dimensional MHD solution of
black hole accretion, with the inclusion of an evenly symmetric
magnetic field (Fig. 1). The steady state and self-similar
assumption in the radial direction are adopted to simplify the
calculation. One caveat in our solution  is that we fail to obtain
the solution in the full $r-\theta$ space. Our integration starts
from the equatorial plane but stops at a certain $\theta_{B}$ angle
since the solution becomes unphysical beyond that angle. But we
believe that the solution obtained is still physical since all the
physical quantities from $\theta=90^{\circ}$ to $\theta_B$ are
physical.

An inflow-outflow solution is found. Around the equatorial plane, it
is the inflow. Above and below the equatorial plane beyond a certain
$\theta$ angle, it is outflow (wind) (Fig. 3). This structure is
same as that obtained in previous numerical simulation works (e.g.,
Yuan et al. 2012; Narayan et al. 2012; Sadowski et al. 2013; Yuan et
al. 2015). We have also calculated some properties of winds and
found good consistency with those obtained in Yuan et al. (2015) and
Yuan et al. (2012), which are based on the MHD numerical simulation
results. For example, we find that the specific angular momentum of
wind is larger than that of inflow thus wind can effectively
transfer angular momentum; the poloidal velocity of wind at both
their origin and infinity calculated in the present analytical work
is also quite similar to those obtained by numerical simulations. At
last, the driving force of wind is analyzed and found to be the sum
of centrifugal force, gradient of gas and magnetic pressure. This is
again consistent with the results obtained in Yuan et al. (2015).

\section*{Acknowledgments}
We thank Zhao-Ming Gan and Mao-chun Wu for useful discussions. This
project was supported in part by the National Basic Research Program
of China (973 Program, grant 2014CB845800), the Strategic Priority
Research Program ¡°The Emergence of Cosmological Structures¡± of CAS
(grant XDB09000000), the Natural Science Foundation of China (grants
11133005 and 11573051), and the CAS/SAFEA International Partnership
Program for Creative Research Teams.


\begin{thebibliography}{99}
\bibitem[\protect\citeauthoryear{Abbassi et al.}{2008}]{Abbassi et al. 2008} Abbassi, S., Ghanbari, J., Najjar, S., 2008, MNRAS, 388, 663
\bibitem[\protect\citeauthoryear{Akizuki \& Fukue}{2006}]{Akizuki and Fukue 2006} Akizuki, C., Fukue, J., 2006, PASJ, 58, 469
\bibitem[\protect\citeauthoryear{Balbus \& Hawley}{1998}]{Balbus and Hawley 1998} Balbus, S. A., \& Hawley, J. F., 1998, Rev. Mod. Phys., 70, 1
\bibitem[\protect\citeauthoryear{Blandford \& Begelman}{1999}]{Blandford and Begelman 1999} Blandford, R. D., Begelman, M. C. 1999, MNRAS, 303, L1
\bibitem[\protect\citeauthoryear{Blandford \& Begelman}{2004}]{Blandford and Begelman 2004} Blandford, R. D., Begelman, M. C. 2004, MNRAS, 349, 68
\bibitem[\protect\citeauthoryear{Blandford \& Payne}{1982}]{Blandford and Payne 1982} Blandford, R., Payne D. G., 1982, MNRAS, 199, 883
\bibitem[\protect\citeauthoryear{Bu et al.}{2009}]{Bu et al. 2009} Bu, D., Yuan, F., Xie, F., 2009, MNRAS, 392, 325
\bibitem[\protect\citeauthoryear{Bu et al.}{2013}]{Bu et al. 2013} Bu, D., Yuan, F., Wu, M., Cuadra, J., 2013, MNRAS, 434, 1692
\bibitem[\protect\citeauthoryear{Cao}{2011}]{Cao 2011} Cao, Xinwu., 2011, ApJ, 737, 94
\bibitem[\protect\citeauthoryear{Gu}{2015}]{Gu 2015} Gu, W. M., 2015, ApJ, 799, 71

\bibitem[\protect\citeauthoryear{Ichimaru}{1977}]{Ichimaru 1977} Ichimaru, S., 1977, ApJ, 214, 840
\bibitem[\protect\citeauthoryear{Igumenshchev \& Abramowicz}{1999}]{Igumenshchev and Abramowicz 1999} Igumenshchev, I. V., Abramowicz M. A., 1999, MNRAS, 303, 309
\bibitem[\protect\citeauthoryear{Igumenshchev \& Abramowicz}{2000}]{Igumenshchev and Abramowicz 2000} Igumenshchev, I. V., Abramowicz M. A., 2000, ApJS, 130, 463
\bibitem[\protect\citeauthoryear{Jiao \& Wu}{2011}]{Jiao and Wu 2011} Jiao, C. L., Wu, X. B., 2011, ApJ, 733, 112
\bibitem[\protect\citeauthoryear{Li \& Begelman}{2014}]{Li and Begelman 2014} Li, S. L., Begelman, M. C., 2014, ApJ, 786, 6
\bibitem[\protect\citeauthoryear{Li, Ostriker \& Sunyaev}{2013}]{Li Ostriker and Sunyaev 2013} Li, J., Ostriker J., Sunyaev R., 2013, ApJ, 767, 105
\bibitem[\protect\citeauthoryear{Lovelace et al.}{1994}]{Lovelace et al. 1994} Lovelace, R. V. E., Romanova, M. M., Newman, W. I., 1994, ApJ, 437, 136
\bibitem[\protect\citeauthoryear{Mosallanezhad et al.}{2014}] {Mosallanezhad et al. 2014} Mosallanezhad A., Abbassi S., Beiranvand N., 2014, MNRAS, 437, 3112
\bibitem[\protect\citeauthoryear{Mou et al.}{2014}]{Mou et al. 2014} Mou, G., Yuan, F., Bu, D., Sun, M., Su, M., 2014, ApJ, 790, 109
\bibitem[\protect\citeauthoryear{Mou et al.}{2015}]{Mou et al. 2015} Mou, G., Yuan, F., Gan, Z., Sun, M. 2015, ApJ, arXiv:1505.00892
\bibitem[\protect\citeauthoryear{Narayan et al.}{2000}]{Narayan et al. 2000} Narayan, R., Igumenshchev, I. V., \& Abramowicz, M. A. 2000, ApJ, 593, 798
\bibitem[\protect\citeauthoryear{Narayan et al.}{2012}]{Narayan et al. 2012} Narayan, R., Sadowski, A., Penna, R. F., et al. 2012, MNRAS, 426, 3241
\bibitem[\protect\citeauthoryear{Narayan \& Yi}{1994}]{Narayan and Yi 1994} Narayan, R., \& Yi, I., 1994, ApJ,428, L13
\bibitem[\protect\citeauthoryear{Narayan \& Yi}{1995}]{Narayan and Yi 1995} Narayan, R., \& Yi, I., 1995, ApJ, 444, 238
\bibitem[\protect\citeauthoryear{Ostriker et al.}{2010}]{Ostriker et al. 2010} Ostriker, J. P., Choi E., Ciotti L., Novak G. S., Proga D., 2010, ApJ, 722, 642
\bibitem[\protect\citeauthoryear{Quataert \& Gruzinov}{2000}]{Quataert and Gruzinov 2000} Quataert, E., \& Gruzinov, A. 2000, ApJ, 539, 809
\bibitem[\protect\citeauthoryear{Rees et al.}{1982}]{Rees et al. 1982} Rees M. J., Begelman M. C., Blandford R. D., Phinney E. S., 1982, Nature, 295, 17
\bibitem[\protect\citeauthoryear{Sadowski et al.}{2013}]{Sadowski et al. 2013} Sadowski, A., Narayan, R., Penna, R., \& Zhu, Y. 2013, MNRAS, 436, 3856
\bibitem[\protect\citeauthoryear{Samadi}{2016}]{Samadi 2016} Samadi M., Abbassi S., 2016, MNRAS, 455
, 3381
\bibitem[\protect\citeauthoryear{Stone, Pringle \& Begelman}{1999}]{Stone Pringle and Begelman 1999} Stone, J. M., Pringle J. E., Begelman M. C., 1999, MNRAS, 310, 1002
\bibitem[\protect\citeauthoryear{Stone \& Pringle}{2001}]{Stone and Pringle 2001} Stone, J. M., Pringle J. E., 2001, MNRAS, 322, 461
\bibitem[\protect\citeauthoryear{Tanaka \& Menou}{2006}]{Tanaka and Menou 2006} Tanaka, T., \& Menou, K. 2006, ApJ, 649, 345
\bibitem[\protect\citeauthoryear{Wang et al.}{2013}]{Wang et al. 2013} Wang, Q. D., Nowak, M. A, Markoff, S. B, et al., 2013, Science, 341, 981
\bibitem[\protect\citeauthoryear{Xu \& Chen}{1997}]{Xu and Chen 1997} Xu, G., Chen, X., 1997, ApJ, 489, L29
\bibitem[\protect\citeauthoryear{Xue \& Wang}{2005}]{Xue and Wang 2005} Xue, L., Wang, J.-C., 2005, ApJ,623, 372
\bibitem[\protect\citeauthoryear{Yuan, Bu \& Wu}{2012a}]{Yuan Bu and Wu 2012} Yuan, F., Bu, D., Wu, M., 2012a, ApJ, 761, 130
\bibitem[\protect\citeauthoryear{Yuan \& Narayan}{2014}]{Yuan and Narayan 2014} Yuan, F., Narayan, R., 2014, ARA\&A, 52, 529
\bibitem[\protect\citeauthoryear{Yuan et al.}{2012b}]{Yuan Wu and Bu 2012} Yuan, F., Wu M., Bu D., 2012b, ApJ, 761, 129
\bibitem[\protect\citeauthoryear{Yuan et al.}{2015}]{Yuan et al.2015} Yuan, F., Gan, Z., Narayan, R., Sadowski, A., Bu, D., Bai, X. 2015, ApJ, 804, 101
\bibitem[\protect\citeauthoryear{Zhang \& Dai}{2008}]{Zhang and Dai 2008} Zhang, D., Dai, Z. G., 2008, MNRAS, 388, 1409




\end{thebibliography}
\end{document}